\documentclass[twocolumn,aps,pra,showpacs, preprintnumbers,letterpaper,amsmath,amssymb,floatfix,superscriptaddress]{revtex4-1}

\usepackage{color}
\usepackage{graphicx}      
\usepackage{subfigure}
\usepackage{dcolumn}
\usepackage{textcomp}
\usepackage{longtable}     
\usepackage{url}           
\usepackage{bm}            
\usepackage{appendix}
\usepackage{hyperref}
\usepackage{soul}

\newcommand{\ket}[1]{\left|{#1}\right\rangle}

\begin{document}

\title{Detection of NaRb Feshbach molecule by photodissociation}

\author{Fan Jia}
\affiliation{Department of Physics, The Chinese University of Hong Kong, Hong Kong, China}
\author{Zhichao Guo}
\affiliation{Department of Physics, The Chinese University of Hong Kong, Hong Kong, China}
\author{Lintao Li}
\affiliation{Department of Physics, The Chinese University of Hong Kong, Hong Kong, China}
\author{Dajun Wang}
\email{djwang@cuhk.edu.hk}
\affiliation{Department of Physics, The Chinese University of Hong Kong, Hong Kong, China}
\affiliation{The Chinese University of Hong Kong Shenzhen Research Institute, Shenzhen, China}

\date{\today}

\begin{abstract}
We demonstrate detection of NaRb Feshbach molecules at high magnetic field by combining molecular photodissociation and absorption imaging of the photofragments.  The photodissociation process is carried out via a spectroscopically selected hyperfine Zeeman level correlated with the Na ($3P_{3/2}$) + Rb ($5S_{1/2}$) asymptote which, following spontaneous emission and optical pumping, leads to ground-state atoms in a single level with near unity probability. Subsequent to the dissociation, the number of molecules is obtained by detecting the resultant $^{23}$Na and $^{87}$Rb atoms. We have also studied the heating effect caused by the photodissociation process and optimized the detection protocol for extracting the temperature of the molecular cloud. This method enables the $in~situ$ detection of fast time scale collision dynamics between NaRb Feshbach molecules and will be a valuable capability in studying few-body physics involving molecules.


\end{abstract}

\maketitle



\section{Introduction}
\label{intro}

Magnetically tunable Feshbach resonances in ultracold atomic gases make it possible to explore a vast range of phenomena~\cite{chin2010feshbach}. Among them, creating weakly bound Feshbach molecules (FMs) by magneto-association is one of the most important as it provides a gateway toward the physics of ultracold molecules~\cite{Kohler06}. Weakly bound heteronuclear FMs can be transferred to the ground state via a two-photon Raman process~\cite{ni2008high,Takekoshi2014,Molony2014,Park2015,Guo2016}, where they have a significant permanent electric dipole moment. This is currently the most successful method for producing ultracold polar molecules for exploring many-body dipolar physics. In addition, FMs have found great applications in few-body and many-body physics, for instance, the Efimov effect~\cite{lompe2010radio,bloom2013tests,klauss2017observation} and the BEC-BCS crossover~\cite{jochim2003bose,greiner2003emergence}.

For all experiments involving FMs, a convenient detection method is crucial for obtaining the basic information, such as number and temperature, of the molecular samples. However, unlike atoms, molecules have much more internal levels, which make it difficult to find cycling transitions necessary for detection with high signal to noise ratios. Therefore, detection of FMs typically relies on dissociating them and then probing via the cycling transition of the resultant atoms. The dissociation can be done by ramping back the magnetic field across the Feshbach resonance~\cite{herbig2003preparation} or by applying an rf pulse resonant with a bound-free transition~\cite{spiegelhalder2010all}. In some special cases, it is also possible to probe weakly bound FMs directly on atomic transitions~\cite{ospelkaus2006ultracold,zirbel2008heteronuclear}. All these methods, however, have some limitations. Ramping the magnetic field takes time and may also introduce extra kinetic energy to the samples~\cite{durr2004dissociation}; rf dissociation and direct imaging on atomic transitions only work for very small binding energies when the bound-free transition strength is still strong enough.

In this paper, we report detection of FMs by first photodissociating them and then detecting the photofragments, all in presence of the high magnetic field. Our method is an extension of the direct FM detection on the atomic transition~\cite{ospelkaus2006ultracold,zirbel2008heteronuclear}, but it works for samples of pure FMs with a large range of binding energies from several kHz to 20 MHz. In this work, after optimizing the laser pulse duration for the photodissociation (PD) and the absorption imaging processes, we can obtain the accurate number and temperature of the NaRb FM sample in about 20 $\mu$s. This method can also be readily adopted for probing dynamics in mixtures of atoms and FMs, e.g., by adding a rf or microwave assisted atom detection first before detecting the FMs. It thus complements the other aforementioned detection methods and may find applications for detecting fast time scale ultracold collision dynamics involving molecules~\cite{wang2019observation}.

The paper is organized as follows. In Sec.~\ref{path}, we discuss the PD pathways for NaRb FMs and the energy level scheme for imaging the resultant atoms in high magnetic field. Sec.~\ref{spect} is devoted to the spectroscopic investigation of the PD pathways. The results of probing the molecules via PD are presented in Sec.~\ref{image} before concluding in Sec.~\ref{conclude}.


\section{The photodissociation pathway}
\label{path}

\begin{figure}
\includegraphics[width=0.8\linewidth]{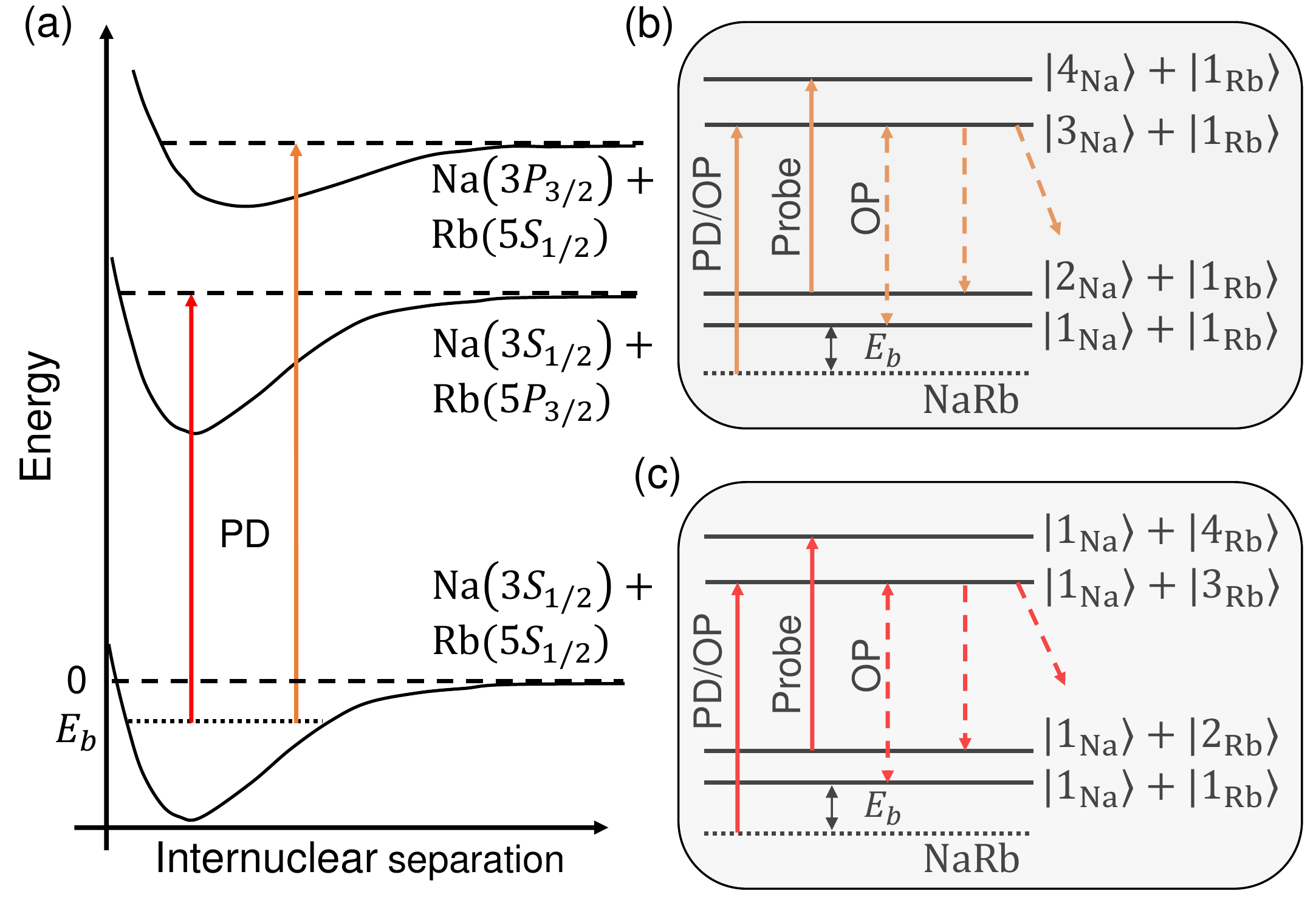}
\caption{Photodissociation pathway and high-field imaging of $^{23}$Na$^{87}$Rb FMs. (a) ground-state NaRb FMs can be dissociated via the $\ket{3_{\rm Na}} + \ket{1_{\rm Rb}} $ or the $\ket{1_{\rm Na}} + \ket{3_{\rm Rb}}$ excited-state atomic limit (see text for detailed symbol designations). Information of the NaRb sample can be obtained by detecting (b) the Na and/or (c) the Rb atoms produced by PD. At high magnetic field, these atoms can be imaged with the closed probe transition after OP. The same OP and probing levels are also used for detecting atomic species directly. The leakage to other levels during the OP is small and can be taken into account to get the molecule number accurately. }
\label{fig1}
\end{figure}

As shown in Fig.~\ref{fig1}, in the ground state, the dissociation asymptote of the FM is $\ket{1_{\rm Na}} + \ket{1_{\rm Rb}} \equiv$ Na$\ket{3S_{1/2}, F = 1, m_F = 1}$ + Rb$\ket{5S_{1/2}, F = 1, m_F = 1}$. Starting from either of the two atomic hyperfine levels, there are no closed transitions for imaging the FMs or even atoms directly. In our experiment, to detect the $\ket{1_{\rm Na}}$ atoms in high magnetic field, we apply an optical pumping (OP) pulse first to transfer them into the $\ket{2_{\rm Na}} \equiv \ket{3S_{1/2}, F = 2, m_F = 2}$ state and then image them with the $\ket{2_{\rm Na}} \rightarrow \ket{4_{\rm Na}}\equiv \ket{3P_{3/2}, m_J^\prime = 3/2, m_I^\prime = 3/2}$ cycling transition. The OP relies on the $\ket{3_{\rm Na}} \equiv \ket{3P_{3/2}, m_J^\prime = 1/2, m_I^\prime = 3/2}$ excited-state hyperfine Zeeman level which decays spontaneously to $\ket{2_{\rm Na}}$ and $\ket{1_{\rm Na}}$ with branching ratios $\eta$ of 66.54\% and 31.13\%, respectively. The leakage to other ground-state levels is 2.3\%. The $\ket{1_{\rm Rb}}$ atoms can also be detected following the same route, with $\ket{2_{\rm Rb}} \equiv \ket{5S_{1/2}, F = 2, m_F = 2}$, $\ket{3_{\rm Rb}} \equiv \ket{5P_{3/2}, m_J^\prime = 1/2, m_I^\prime = 3/2}$, and $\ket{4_{\rm Rb}}\equiv \ket{5P_{3/2}, m_J^\prime = 3/2, m_I^\prime = 3/2}$. The branching ratios from $\ket{3_{\rm Rb}}$ to $\ket{2_{\rm Rb}}$ and $\ket{1_{\rm Rb}}$ are 64.19\% and 35.13\%, respectively. The leakage to other ground-state levels is only 0.68\%. Here, we use quantum numbers $F$ and $m_F$ to label the ground-state hyperfine Zeeman levels. However, under the high magnetic field, they are not good quantum number in the excited state. Thus $m_J^\prime$ and $m_I^\prime$, which are projections of the electron's total angular momentum and the nuclear spin onto the magnetic field, are used.      

For FM detection, the OP for atom is replaced by PD plus OP (PD/OP). As shown in Fig.~\ref{fig1}, in principle the PD can occur via either the $ \ket{3_{\rm Na}} +\ket{1_{\rm Rb}} $ or the $\ket{1_{\rm Na}} + \ket{3_{\rm Rb}}$ excited-state atomic limits. Starting from FMs, the two atomic limits can be reached by adding the molecule binding energy $E_{\rm b}$ to the $\ket{1_{\rm Na}} \rightarrow \ket{3_{\rm Na}}$ or the $\ket{1_{\rm Rb}} \rightarrow \ket{3_{\rm Rb}}$ OP transition. As these transitions are electric dipole (E1) transitions, the PD process is much more efficient than the rf dissociation which relies on much weaker magnetic dipole (M1) transitions. The photodissociated atom pair in the $ \ket{3_{\rm Na}} + \ket{1_{\rm Rb}} $ ($\ket{1_{\rm Na}} + \ket{3_{\rm Rb}}$) asymptote will result in a $\ket{1_{\rm Rb}}$ ($\ket{1_{\rm Na}}$) atom and a $\ket{2_{\rm Na}}$ ($\ket{2_{\rm Rb}}$) or $\ket{1_{\rm Na}}$ ($\ket{1_{\rm Rb}}$) atom following the spontaneous emission. The $\ket{1_{\rm Na}}$ ($\ket{1_{\rm Rb}}$) atoms are then further optically pumped to the $\ket{2_{\rm Na}}$ ($\ket{2_{\rm Rb}}$) off-resonantly by the PD/OP light in which they can be probed together with atoms already decayed to this state. The other $\ket{1_{\rm Rb}}$ ($\ket{1_{\rm Na}}$) atoms can also be detected following the aforementioned high-field imaging procedure for atoms with the help of an additional OP pulse. Information of the FM sample can be extracted from one or both images.

An important issue to consider in selecting the excited state for PD is molecular levels near the dissociation limit. The existence of such a weakly-bound molecular level, if also reachable by the PD light, will compromise the PD process as the bound-bound transition has a more favorable transition strength and is thus more likely to happen than the bound-free transition. In fact, as will be presented in the next section, such a NaRb level is observed near the $\ket{1_{\rm Na}} + \ket{3_{\rm Rb}}$ asymptote, which excludes usage of this PD pathway for detecting NaRb FMs.


\section{Photodissociation spectroscopy}
\label{spect}


\subsection{Experiment setup}

The experimental system for creating the sample of pure FMs has been discussed in detail before~\cite{wang2015formation,Guo2016}. Briefly, we use the 347.64~G Feshbach resonance between $\ket{1_{\rm Na}}$ and $\ket{1_{\rm Rb}}$ atoms to produce NaRb FMs via magnetoassociation. The only difference is that a pancake shaped optical dipole trap formed by crossing two 946 nm laser beams is used. In this so called ``magic wavelength'' optical trap~\cite{safronova2006frequency}, the Na and Rb samples feel the same trapping frequencies and have overlapping centers of mass which is advantageous for creating FMs. In the final configuration, the measured trap frequencies are ($\omega_x$, $\omega_y$, $\omega_z$) = 2$\pi \times$(40, 40, 193) Hz.  After the magnetic field ramp for magnetoassociation, a strong magnetic field gradient pulse is applied to remove all residue atoms at 335.3 G. Following this procedure, we can obtain typically 7000 NaRb FMs with no detectable residue Na and Rb atoms. The trap lifetime of the molecule sample is more than 20 ms, enough for the subsequent investigations.

In previous works~\cite{wang2015formation,Guo2016}, to detect the FMs, we first dissociate them by ramping the magnetic field reversely across the Feshbach resonance. The high magnetic field is then turned off abruptly and the resultant Na and Rb atoms are detected with standard low-field absorption imaging method.

For the high-field imaging with PD, the $\ket{1_{\rm Rb}} \rightarrow \ket{3_{\rm Rb}}$ PD/OP transition is driven by a DFB laser offset locked~\cite{Schunemann1999offset} to the Rb repumping light which is on resonance with the $\ket{F = 1} \rightarrow \ket{F^\prime = 2}$ transition at zero magnetic field. The laser frequency can be easily tuned by several GHz for compensating the Zeeman shift. The $\ket{1_{\rm Na}} \rightarrow \ket{3_{\rm Na}}$ PD/OP light is derived by frequency shifting the $\ket{F = 1} \rightarrow \ket{F^\prime = 2}$ Na cooling light by $\sim+700$ MHz. This is accomplished with an acousto-optical modulator (AOM) in the double-pass configuration. Similarly, the $\ket{2_{\rm Na}} \rightarrow \ket{4_{\rm Na}}$ and $\ket{2_{\rm Rb}} \rightarrow \ket{4_{\rm Rb}}$ probe frequencies are also obtained with AOMs in double-pass configurations.

The PD/OP light propagates along the magnetic field direction and is $\sigma^+$ polarized. To ensure a homogeneous illumination, the PD/OP light has a large beam size of 1.1 mm which is more than 50 times larger than the typical size of the FM sample. The probe light is linear polarized with both its propagation direction and polarization perpendicular to the magnetic field. It can thus drive the $\sigma^+$ imaging transition with half of the maximum absorption cross section. We use two individually controlled CCD cameras, one for each species, to detect the fragmented Na and Rb atoms. In a single run, we can obtain absorption images of both species.


\subsection{Photodissociation via the $\ket{1_{\rm Na}} + \ket{3_{\rm Rb}}$ limit}

To probe the PD transition near the $\ket{1_{\rm Na}} + \ket{3_{\rm Rb}}$ limit, we start from a pure NaRb FM sample at 335.3 G. The magnetic field is then ramped to a desired value in 4 ms. After another 5 ms for the magnetic field to stabilize, a 110 $\mu$W PD/OP pulse is applied for 20 $\mu$s to dissociate the molecules and optical pump the resultant Rb atoms to the $\ket{2_{\rm Rb}}$ state. Subsequently, these atoms are detected with a 50~$\mu$s high-field imaging pulse. In addition, we also detect the remaining molecules with the low-field imaging method after finishing the high-field imaging. The same procedure is then repeated after stepping the PD/OP light frequency for obtaining the PD spectrum at the selected magnetic field.

\begin{figure}[h!]
\centering
\includegraphics[width= 0.85 \linewidth]{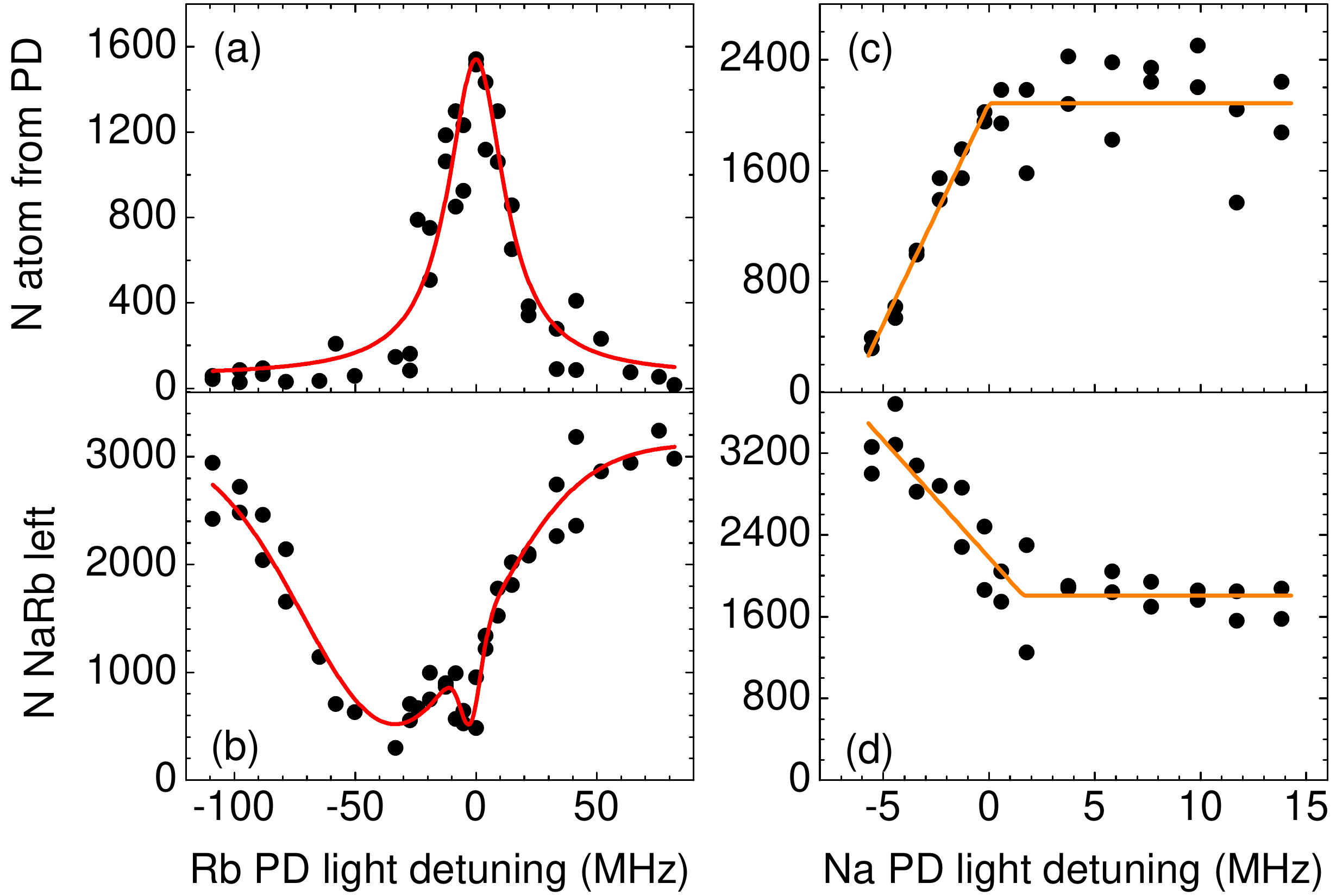}
\caption{Photodissociation spectrum via different pathways. (a) Rb photofragment atoms detected by high-field (343.2 G) imaging and (b) remaining molecules detected by the low-field imaging method after PD via the $\ket{1_{\rm Na}} + \ket{3_{\rm Rb}}$ limit. Besides the PD resonance at 0 MHz, an additional feature from a near threshold bound state is also observed at around -35 MHz. The solid curves are from single and double-peak Lorentzian fittings for extracting the line centers. (c) Na photofragment atoms detected by high-field (335.3 G) imaging and (d) remaining molecules detected by the low-field imaging method after PD via the $\ket{3_{\rm Na}} + \ket{1_{\rm Rb}}$ limit. The solid curves are from piecewise fitting for estimating the PD threshold. }
\label{fig10}
\end{figure}

An example PD spectrum obtained via the high-field imaging method is shown in Fig.~\ref{fig10}(a). As the NaRb molecular potentials correlated with the $\ket{1_{\rm Na}} + \ket{3_{\rm Rb}}$ limit are all attractive, a plateau is expected for the above threshold portion of the PD lineshape. Contrary to this expectation, we observe a nearly symmetric lineshape. This is caused by a near threshold NaRb bound state which is revealed by detecting the remaining molecules with the low-field imaging method. As shown in Fig.~\ref{fig10}(b), the binding energy of this excited NaRb state is only 35 MHz. As the bound-bound transition strength is much stronger, with the same PD/OP power, a very broad bound-bound transition linewidth is observed. This makes it essentially impossible to drive a pure bound-free PD transition. Bound excited NaRb state can decay spontaneously to ground NaRb bound states which are dark to the PD/OP and the imaging light. The inevitable bound-bound transition here thus leads to a significant undercount of the molecule number in the high-field imaging signal. As a result, this PD path is not suitable for detecting NaRb FMs.

\subsection{Photodissociation via the $\ket{3_{\rm Na}} + \ket{1_{\rm Rb}}$ limit}

A similar procedure is used to obtain the PD spectrum near the $\ket{3_{\rm Na}} + \ket{1_{\rm Rb}}$ limit. Fig.~\ref{fig10}(c) shows an example spectrum obtained by detecting the resultant Na atoms at high magnetic field following a 100 $\mu$W, 5 $\mu$s PD/OP pulse. The ``shelf'' like lineshape with an above threshold plateau is typical for molecular PD~\cite{mcguyer2013nonadiabatic}. Unlike that with the $\ket{1_{\rm Na}} + \ket{3_{\rm Rb}}$ limit, the remaining molecules detected with low-field imaging method in Fig.~\ref{fig10}(d) mirrors the Na fragment in Fig.~\ref{fig10}(c) with no additional spectroscopic features. In addition, the sum of the numbers obtained in Fig.~\ref{fig10}(c) and Fig.~\ref{fig10}(d) is constant and equals to the original molecule number measured with the low-field imaging method without the PD. This suggests that the PD process is clear and all the dissociated molecules are converted to $\ket{2_{\rm Na}}$ and $\ket{1_{\rm Rb}}$ atoms. This PD pathway is thus selected for the high-field FM detection.   

The PD spectrum in Fig.~\ref{fig10}(c) is obtained at 335.3 G where the FM is bound by 22.38 MHz~\cite{wang2015formation}. Thanks to the strong E1 coupling, PD still occurs efficiently with rather low PD/OP light power. However, at this binding energy, the dissociation by rf or microwave driven M1 transitions is practically impossible due to the small bound-to-free wavefunction overlap. In fact, we have observed that rf dissociation already becomes difficult for binding energies large than 2 MHz even with very high rf powers.

\section{High magnetic field FM detection}\label{image}

\begin{figure}
\centering
\includegraphics[width = 0.8 \linewidth]{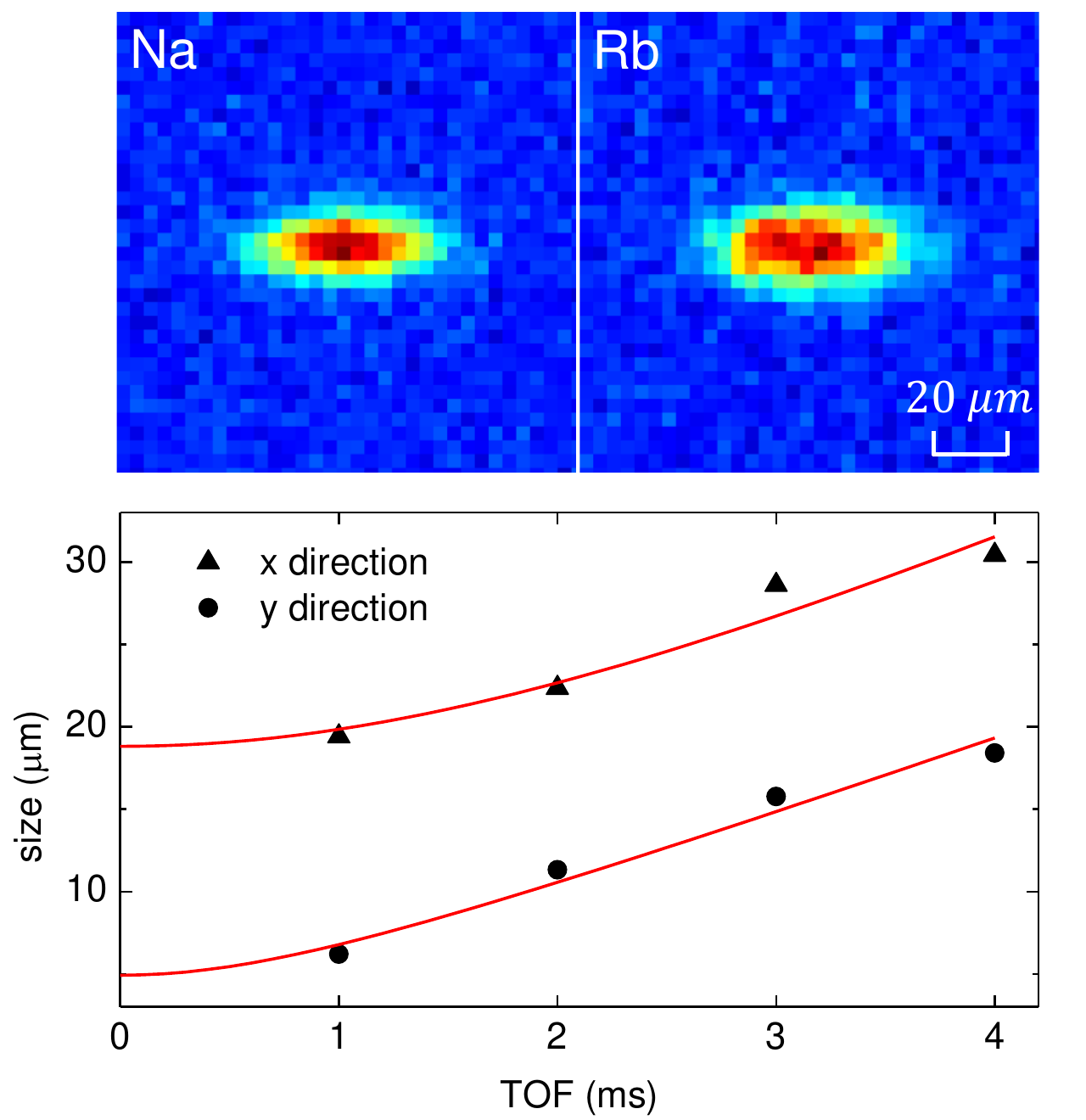}
 \caption{Detection of NaRb FMs by PD. Top: high-field absorption images of the FMs by probing the Na (left) and Rb (right) photofragments. Bottom: x and y sizes of the FM sample following TOF expansion before the PD. The red curves are fits to Eq.~(\ref{eq1}) for extracting the temperature. }
\label{fig3}
\end{figure}

Figure~\ref{fig3} shows single-shot absorption images of the Na and Rb fragments taken at 343.2 G following the PD/OP pulse. While Na atoms are pumped to $\ket{2_{\rm Na}}$ level and can be imaged immediately, an additional OP pulse is needed for imaging Rb atoms which are in the $\ket{1_{\rm Rb}}$ level after the PD. To ensure a complete PD, the PD/OP light frequency is tuned near the dissociation limit. Typical for PD processes, the excess photon energy will be converted into kinetic energies of the photofragments. In the current experiment, as the nature linewidth of the $\ket{3_{\rm Na}}$ level is 10 MHz and the linewidth of the PD/OP light is sub-MHz, the fragmented Na and Rb atoms can be easily heated up by hundreds of $\mu$K (1 MHz corresponds to 48~$\mu$K). 

\begin{figure}
\centering
\includegraphics[width = 0.8 \linewidth]{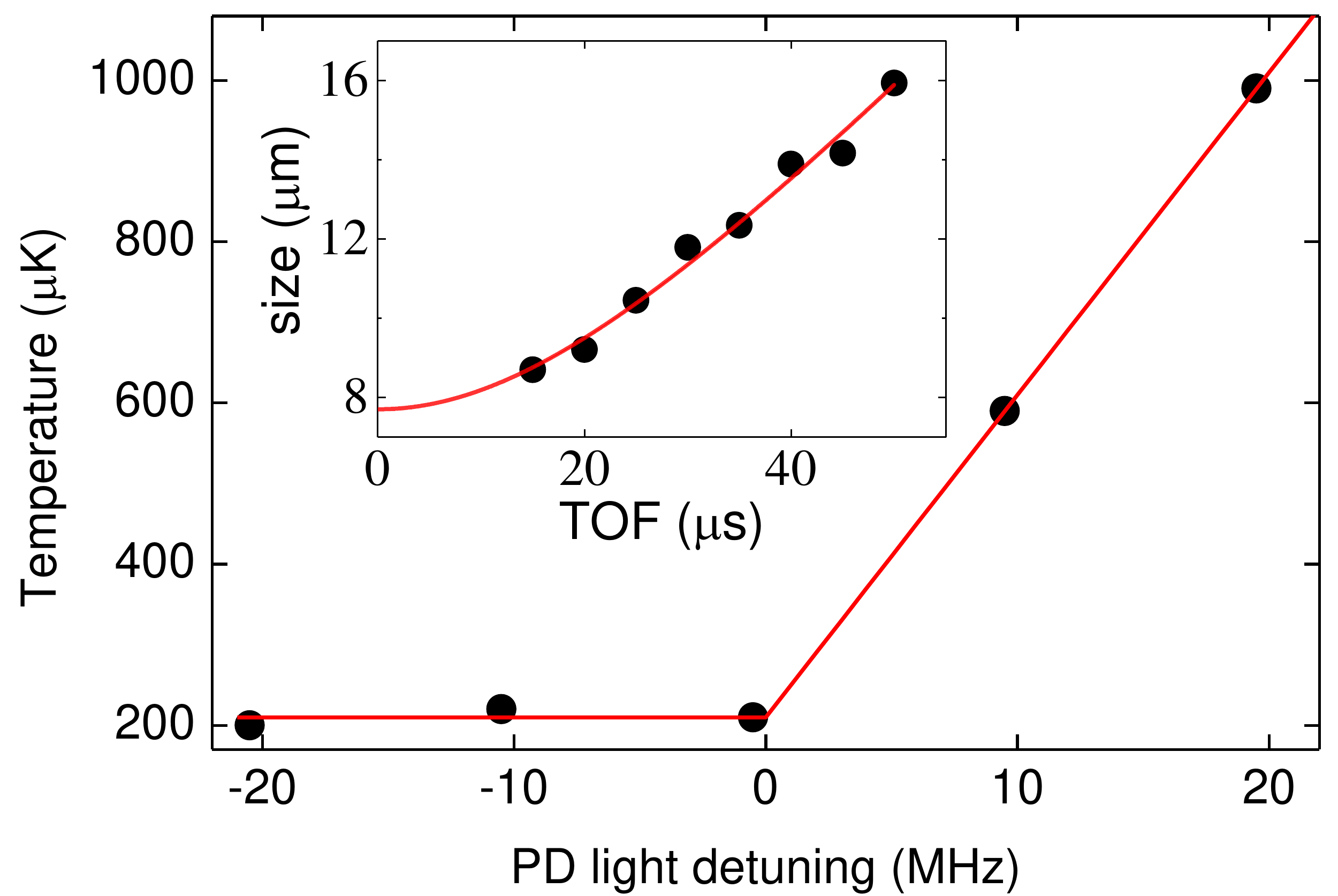}
\caption{The strong heating effect from PD. Temperatures at several PD detunings are obtained from measuring the TOF expansion after the PD. The solid curve is from a piecewise fitting for extracting the PD threshold and the heating rate for above threshold PD. Inset: an example TOF measurement. Very short expansion time is used due to the high temperature. The curve is the fit to Eq.~(\ref{eq1}) for obtaining the temperature.}
\label{fig4}
\end{figure}

This heating effect is demonstrated in Fig.~\ref{fig4} in which the temperature is obtained from measuring the time-of-flight (TOF) expansion (inset) of the Na cloud after the PD/OP pulse. Fit the measured size to the TOF expansion formula for thermal gases
\begin{equation}
\sigma(t)^2=\sigma_0^2+ \frac{k_B T t^2}{M},
\label{eq1}
\end{equation}
we obtain temperatures $T$ for several PD/OP light frequencies. Here $k_B$ is the Boltzmann constant, $\sigma_0$ is the in trap size, and $M$ is the mass of the Na atom. For the above threshold portion, a heating rate of 40(2)~$\mu$K/MHz is obtained from the piecewise fitting. If we ignore the small initial NaRb sample temperature, the heating rate calculated from momentum and energy conversations is $38~\mu$K/MHz for the Na fragments. This is in good agreement with the observation. For Rb, a smaller heating rate of $10~\mu$K/MHz is expected. As shown in Fig.~\ref{fig4}, we also observe that the Na cloud is heated up to about 210~$\mu$K even with PD/OP light frequency below the dissociation threshold. We believe this is due to power broadening and the nature linewidth of the PD/OP transition.

Since the temperatures of the photofragments are much higher than the typical trap depth, the Na and Rb clouds will expand quickly with or without the trap. The slightly larger size of the Rb cloud in Fig.~\ref{fig3} is a result of this expansion as the Rb image is taken $50$~$\mu$s after the Na image due to the need for the additional OP. To get the correct number of FMs from the images, it is important to ensure a complete PD of all FMs. Meanwhile, the PD induced heating also makes it critical to use a short duration to avoid number loss and large size expansion before the detection. To this end, the PD/OP pulse power is increased to 2 mW which is strong enough to dissociate all the molecules in 3~$\mu$s. At the same experimental condition for obtaining Fig.~\ref{fig3}, the molecule numbers extracted from the Na and Rb images after averaging 10 shots are $6740(280)$ and $6190(360)$, respectively. The small corrections due to the leakages in the OP are already taken into account. The two numbers, which should both equal to the number of molecules and thus identical, are slightly different. The same issue has also been observed in the low-field imaging method~\cite{ye2018collisions,Guo2018} and can be eliminated by a more careful number calibration. Within the measurement uncertainties, the molecule numbers measured by the PD method agree with those obtained from the magneto-dissociation and low-field imaging method.

The strong heating effect of the PD also complicates the temperature measurement. To mitigate this problem, the detection is performed on the Rb cloud due to its lower sample temperature than that of the Na cloud. In addition, we also set the PD/OP light frequency about 1 MHz below the PD threshold to reduce the heating as much as possible without sacrificing the PD efficiency. To measure the sample temperature, the NaRb FMs are first released from the optical trap. After various TOF, they are then photodissociated by the 3~$\mu$s PD/OP pulse. An 8~$\mu$s Rb OP pulse is applied immediately afterwards to pump the $\ket{1_{\rm Rb}}$ fragments to $\ket{2_{\rm Rb}}$ which are then imaged by a 10~$\mu$s high-field image pulse. During the whole PD and imaging procedure, the size of the Rb fragment expands about 1 $\mu$m, which is less than 20\% of $\sigma_0$ of the FMs. The bottom panel in Fig.~\ref{fig3} shows the measured Rb sizes versus the expansion time. Fit the data to Eq.~(\ref{eq1}) by replacing $M$ with the mass of the NaRb molecule, we obtain a molecule sample temperature $T = 390(40)$~nK.

Since the TOF expansion here occurs before the PD, the PD induced heating affects mainly the measured $\sigma_0$. In principle, longer TOFs can be used to reduce this effect. When $\sigma(t)$ becomes much large than $\sigma_0$, the temperature can be extracted directly from the slope of the expansion curve without involving $\sigma_0$. In our experiment, however, the TOF is limited to several ms due to the small molecule number. Nevertheless, under the same sample condition, the temperature measured in the PD assisted high-field imaging method already agrees within less than 5\% with that obtained using the low-field imaging method. We thus conclude that both the number and temperature of the FMs can be accurately obtained with the PD assisted high-field imaging method.


\section{Conclusion}
\label{conclude}

To summarize, we have studied the PD of NaRb FMs via the Na($3P_{3/2}$) + Rb($5S_{1/2}$) and the Na($3S_{1/2}$) + Rb($5P_{3/2}$) excited-state dissociation limits and selected the Na$\ket{3P_{3/2}, m_J^\prime = 1/2, m_I^\prime = 3/2}$ hyperfine Zeeman level of the former path for detecting the FMs. We investigated the heating effect accompanying the PD process and developed a protocol which has allowed us to obtain the accurate number and temperature of the molecular sample. Beside its capability of detecting pure FMs with both small and large binding energies, this detection method is also useful for studying collisions between atom and FMs. However, as relatively strong intensities are needed for the PD, this scheme cannot directly distinguish atoms and FMs prepared in the same threshold when the FM binding energies are smaller than the nature linewidth. Fortunately, this issue can be solved by the addition of a high-resolution rf or microwave pulse to transfer the atoms to the right energy level and then probed by the cycling transition. After the atoms are removed by the probe light, the FMs can then be detected following our protocol. In addition, combined with the stimulated Raman process, this method can also be readily adopted for detection of ground-state molecules.

\begin{acknowledgments}

This work is supported by Hong Kong RGC General Research Fund (grants 14301815 and 14303317) and the Collaborative Research Fund C6026-16W.

\end{acknowledgments}


\begin{thebibliography}{24}%
\makeatletter
\providecommand \@ifxundefined [1]{%
 \@ifx{#1\undefined}
}%
\providecommand \@ifnum [1]{%
 \ifnum #1\expandafter \@firstoftwo
 \else \expandafter \@secondoftwo
 \fi
}%
\providecommand \@ifx [1]{%
 \ifx #1\expandafter \@firstoftwo
 \else \expandafter \@secondoftwo
 \fi
}%
\providecommand \natexlab [1]{#1}%
\providecommand \enquote  [1]{``#1''}%
\providecommand \bibnamefont  [1]{#1}%
\providecommand \bibfnamefont [1]{#1}%
\providecommand \citenamefont [1]{#1}%
\providecommand \href@noop [0]{\@secondoftwo}%
\providecommand \href [0]{\begingroup \@sanitize@url \@href}%
\providecommand \@href[1]{\@@startlink{#1}\@@href}%
\providecommand \@@href[1]{\endgroup#1\@@endlink}%
\providecommand \@sanitize@url [0]{\catcode `\\12\catcode `\$12\catcode
  `\&12\catcode `\#12\catcode `\^12\catcode `\_12\catcode `\%12\relax}%
\providecommand \@@startlink[1]{}%
\providecommand \@@endlink[0]{}%
\providecommand \url  [0]{\begingroup\@sanitize@url \@url }%
\providecommand \@url [1]{\endgroup\@href {#1}{\urlprefix }}%
\providecommand \urlprefix  [0]{URL }%
\providecommand \Eprint [0]{\href }%
\providecommand \doibase [0]{http://dx.doi.org/}%
\providecommand \selectlanguage [0]{\@gobble}%
\providecommand \bibinfo  [0]{\@secondoftwo}%
\providecommand \bibfield  [0]{\@secondoftwo}%
\providecommand \translation [1]{[#1]}%
\providecommand \BibitemOpen [0]{}%
\providecommand \bibitemStop [0]{}%
\providecommand \bibitemNoStop [0]{.\EOS\space}%
\providecommand \EOS [0]{\spacefactor3000\relax}%
\providecommand \BibitemShut  [1]{\csname bibitem#1\endcsname}%
\let\auto@bib@innerbib\@empty
\bibitem [{\citenamefont {Chin}\ \emph {et~al.}(2010)\citenamefont {Chin},
  \citenamefont {Grimm}, \citenamefont {Julienne},\ and\ \citenamefont
  {Tiesinga}}]{chin2010feshbach}%
  \BibitemOpen
  \bibfield  {author} {\bibinfo {author} {\bibfnamefont {C.}~\bibnamefont
  {Chin}}, \bibinfo {author} {\bibfnamefont {R.}~\bibnamefont {Grimm}},
  \bibinfo {author} {\bibfnamefont {P.}~\bibnamefont {Julienne}}, \ and\
  \bibinfo {author} {\bibfnamefont {E.}~\bibnamefont {Tiesinga}},\ }\href@noop
  {} {\bibfield  {journal} {\bibinfo  {journal} {Rev. Mod. Phys.}\ }\textbf
  {\bibinfo {volume} {82}},\ \bibinfo {pages} {1225} (\bibinfo {year}
  {2010})}\BibitemShut {NoStop}%
\bibitem [{\citenamefont {K\"{o}hler}\ \emph {et~al.}(2006)\citenamefont
  {K\"{o}hler}, \citenamefont {G\'{o}ral},\ and\ \citenamefont
  {Julienne}}]{Kohler06}%
  \BibitemOpen
  \bibfield  {author} {\bibinfo {author} {\bibfnamefont {T.}~\bibnamefont
  {K\"{o}hler}}, \bibinfo {author} {\bibfnamefont {K.}~\bibnamefont
  {G\'{o}ral}}, \ and\ \bibinfo {author} {\bibfnamefont {P.~S.}\ \bibnamefont
  {Julienne}},\ }\href@noop {} {\bibfield  {journal} {\bibinfo  {journal} {Rev.
  Mod. Phys.}\ }\textbf {\bibinfo {volume} {78}},\ \bibinfo {eid} {1311}
  (\bibinfo {year} {2006})}\BibitemShut {NoStop}%
\bibitem [{\citenamefont {Ni}\ \emph {et~al.}(2008)\citenamefont {Ni},
  \citenamefont {Ospelkaus}, \citenamefont {De~Miranda}, \citenamefont {Pe'Er},
  \citenamefont {Neyenhuis}, \citenamefont {Zirbel}, \citenamefont
  {Kotochigova}, \citenamefont {Julienne}, \citenamefont {Jin},\ and\
  \citenamefont {Ye}}]{ni2008high}%
  \BibitemOpen
  \bibfield  {author} {\bibinfo {author} {\bibfnamefont {K.-K.}\ \bibnamefont
  {Ni}}, \bibinfo {author} {\bibfnamefont {S.}~\bibnamefont {Ospelkaus}},
  \bibinfo {author} {\bibfnamefont {M.}~\bibnamefont {De~Miranda}}, \bibinfo
  {author} {\bibfnamefont {A.}~\bibnamefont {Pe'Er}}, \bibinfo {author}
  {\bibfnamefont {B.}~\bibnamefont {Neyenhuis}}, \bibinfo {author}
  {\bibfnamefont {J.}~\bibnamefont {Zirbel}}, \bibinfo {author} {\bibfnamefont
  {S.}~\bibnamefont {Kotochigova}}, \bibinfo {author} {\bibfnamefont
  {P.}~\bibnamefont {Julienne}}, \bibinfo {author} {\bibfnamefont
  {D.}~\bibnamefont {Jin}}, \ and\ \bibinfo {author} {\bibfnamefont
  {J.}~\bibnamefont {Ye}},\ }\href@noop {} {\bibfield  {journal} {\bibinfo
  {journal} {science}\ }\textbf {\bibinfo {volume} {322}},\ \bibinfo {pages}
  {231} (\bibinfo {year} {2008})}\BibitemShut {NoStop}%
\bibitem [{\citenamefont {Takekoshi}\ \emph {et~al.}(2014)\citenamefont
  {Takekoshi}, \citenamefont {Reichs\"ollner}, \citenamefont {Schindewolf},
  \citenamefont {Hutson}, \citenamefont {Le~Sueur}, \citenamefont {Dulieu},
  \citenamefont {Ferlaino}, \citenamefont {Grimm},\ and\ \citenamefont
  {N\"agerl}}]{Takekoshi2014}%
  \BibitemOpen
  \bibfield  {author} {\bibinfo {author} {\bibfnamefont {T.}~\bibnamefont
  {Takekoshi}}, \bibinfo {author} {\bibfnamefont {L.}~\bibnamefont
  {Reichs\"ollner}}, \bibinfo {author} {\bibfnamefont {A.}~\bibnamefont
  {Schindewolf}}, \bibinfo {author} {\bibfnamefont {J.~M.}\ \bibnamefont
  {Hutson}}, \bibinfo {author} {\bibfnamefont {C.~R.}\ \bibnamefont
  {Le~Sueur}}, \bibinfo {author} {\bibfnamefont {O.}~\bibnamefont {Dulieu}},
  \bibinfo {author} {\bibfnamefont {F.}~\bibnamefont {Ferlaino}}, \bibinfo
  {author} {\bibfnamefont {R.}~\bibnamefont {Grimm}}, \ and\ \bibinfo {author}
  {\bibfnamefont {H.-C.}\ \bibnamefont {N\"agerl}},\ }\href {\doibase
  10.1103/PhysRevLett.113.205301} {\bibfield  {journal} {\bibinfo  {journal}
  {Phys. Rev. Lett.}\ }\textbf {\bibinfo {volume} {113}},\ \bibinfo {pages}
  {205301} (\bibinfo {year} {2014})}\BibitemShut {NoStop}%
\bibitem [{\citenamefont {Molony}\ \emph {et~al.}(2014)\citenamefont {Molony},
  \citenamefont {Gregory}, \citenamefont {Ji}, \citenamefont {Lu},
  \citenamefont {K\"oppinger}, \citenamefont {Le~Sueur}, \citenamefont
  {Blackley}, \citenamefont {Hutson},\ and\ \citenamefont
  {Cornish}}]{Molony2014}%
  \BibitemOpen
  \bibfield  {author} {\bibinfo {author} {\bibfnamefont {P.~K.}\ \bibnamefont
  {Molony}}, \bibinfo {author} {\bibfnamefont {P.~D.}\ \bibnamefont {Gregory}},
  \bibinfo {author} {\bibfnamefont {Z.}~\bibnamefont {Ji}}, \bibinfo {author}
  {\bibfnamefont {B.}~\bibnamefont {Lu}}, \bibinfo {author} {\bibfnamefont
  {M.~P.}\ \bibnamefont {K\"oppinger}}, \bibinfo {author} {\bibfnamefont
  {C.~R.}\ \bibnamefont {Le~Sueur}}, \bibinfo {author} {\bibfnamefont {C.~L.}\
  \bibnamefont {Blackley}}, \bibinfo {author} {\bibfnamefont {J.~M.}\
  \bibnamefont {Hutson}}, \ and\ \bibinfo {author} {\bibfnamefont {S.~L.}\
  \bibnamefont {Cornish}},\ }\href {\doibase 10.1103/PhysRevLett.113.255301}
  {\bibfield  {journal} {\bibinfo  {journal} {Phys. Rev. Lett.}\ }\textbf
  {\bibinfo {volume} {113}},\ \bibinfo {pages} {255301} (\bibinfo {year}
  {2014})}\BibitemShut {NoStop}%
\bibitem [{\citenamefont {Park}\ \emph {et~al.}(2015)\citenamefont {Park},
  \citenamefont {Will},\ and\ \citenamefont {Zwierlein}}]{Park2015}%
  \BibitemOpen
  \bibfield  {author} {\bibinfo {author} {\bibfnamefont {J.~W.}\ \bibnamefont
  {Park}}, \bibinfo {author} {\bibfnamefont {S.~A.}\ \bibnamefont {Will}}, \
  and\ \bibinfo {author} {\bibfnamefont {M.~W.}\ \bibnamefont {Zwierlein}},\
  }\href {\doibase 10.1103/PhysRevLett.114.205302} {\bibfield  {journal}
  {\bibinfo  {journal} {Phys. Rev. Lett.}\ }\textbf {\bibinfo {volume} {114}},\
  \bibinfo {pages} {205302} (\bibinfo {year} {2015})}\BibitemShut {NoStop}%
\bibitem [{\citenamefont {Guo}\ \emph {et~al.}(2016)\citenamefont {Guo},
  \citenamefont {Zhu}, \citenamefont {Lu}, \citenamefont {Ye}, \citenamefont
  {Wang}, \citenamefont {Vexiau}, \citenamefont {Bouloufa-Maafa}, \citenamefont
  {Qu\'em\'ener}, \citenamefont {Dulieu},\ and\ \citenamefont
  {Wang}}]{Guo2016}%
  \BibitemOpen
  \bibfield  {author} {\bibinfo {author} {\bibfnamefont {M.}~\bibnamefont
  {Guo}}, \bibinfo {author} {\bibfnamefont {B.}~\bibnamefont {Zhu}}, \bibinfo
  {author} {\bibfnamefont {B.}~\bibnamefont {Lu}}, \bibinfo {author}
  {\bibfnamefont {X.}~\bibnamefont {Ye}}, \bibinfo {author} {\bibfnamefont
  {F.}~\bibnamefont {Wang}}, \bibinfo {author} {\bibfnamefont {R.}~\bibnamefont
  {Vexiau}}, \bibinfo {author} {\bibfnamefont {N.}~\bibnamefont
  {Bouloufa-Maafa}}, \bibinfo {author} {\bibfnamefont {G.}~\bibnamefont
  {Qu\'em\'ener}}, \bibinfo {author} {\bibfnamefont {O.}~\bibnamefont
  {Dulieu}}, \ and\ \bibinfo {author} {\bibfnamefont {D.}~\bibnamefont
  {Wang}},\ }\href {\doibase 10.1103/PhysRevLett.116.205303} {\bibfield
  {journal} {\bibinfo  {journal} {Phys. Rev. Lett.}\ }\textbf {\bibinfo
  {volume} {116}},\ \bibinfo {pages} {205303} (\bibinfo {year}
  {2016})}\BibitemShut {NoStop}%
\bibitem [{\citenamefont {Lompe}\ \emph {et~al.}(2010)\citenamefont {Lompe},
  \citenamefont {Ottenstein}, \citenamefont {Serwane}, \citenamefont {Wenz},
  \citenamefont {Z{\"u}rn},\ and\ \citenamefont {Jochim}}]{lompe2010radio}%
  \BibitemOpen
  \bibfield  {author} {\bibinfo {author} {\bibfnamefont {T.}~\bibnamefont
  {Lompe}}, \bibinfo {author} {\bibfnamefont {T.~B.}\ \bibnamefont
  {Ottenstein}}, \bibinfo {author} {\bibfnamefont {F.}~\bibnamefont {Serwane}},
  \bibinfo {author} {\bibfnamefont {A.~N.}\ \bibnamefont {Wenz}}, \bibinfo
  {author} {\bibfnamefont {G.}~\bibnamefont {Z{\"u}rn}}, \ and\ \bibinfo
  {author} {\bibfnamefont {S.}~\bibnamefont {Jochim}},\ }\href@noop {}
  {\bibfield  {journal} {\bibinfo  {journal} {Science}\ }\textbf {\bibinfo
  {volume} {330}},\ \bibinfo {pages} {940} (\bibinfo {year}
  {2010})}\BibitemShut {NoStop}%
\bibitem [{\citenamefont {Bloom}\ \emph {et~al.}(2013)\citenamefont {Bloom},
  \citenamefont {Hu}, \citenamefont {Cumby},\ and\ \citenamefont
  {Jin}}]{bloom2013tests}%
  \BibitemOpen
  \bibfield  {author} {\bibinfo {author} {\bibfnamefont {R.~S.}\ \bibnamefont
  {Bloom}}, \bibinfo {author} {\bibfnamefont {M.-G.}\ \bibnamefont {Hu}},
  \bibinfo {author} {\bibfnamefont {T.~D.}\ \bibnamefont {Cumby}}, \ and\
  \bibinfo {author} {\bibfnamefont {D.~S.}\ \bibnamefont {Jin}},\ }\href@noop
  {} {\bibfield  {journal} {\bibinfo  {journal} {Phys. Rev. Lett.}\ }\textbf
  {\bibinfo {volume} {111}},\ \bibinfo {pages} {105301} (\bibinfo {year}
  {2013})}\BibitemShut {NoStop}%
\bibitem [{\citenamefont {Klauss}\ \emph {et~al.}(2017)\citenamefont {Klauss},
  \citenamefont {Xie}, \citenamefont {Lopez-Abadia}, \citenamefont {D’Incao},
  \citenamefont {Hadzibabic}, \citenamefont {Jin},\ and\ \citenamefont
  {Cornell}}]{klauss2017observation}%
  \BibitemOpen
  \bibfield  {author} {\bibinfo {author} {\bibfnamefont {C.~E.}\ \bibnamefont
  {Klauss}}, \bibinfo {author} {\bibfnamefont {X.}~\bibnamefont {Xie}},
  \bibinfo {author} {\bibfnamefont {C.}~\bibnamefont {Lopez-Abadia}}, \bibinfo
  {author} {\bibfnamefont {J.~P.}\ \bibnamefont {D’Incao}}, \bibinfo {author}
  {\bibfnamefont {Z.}~\bibnamefont {Hadzibabic}}, \bibinfo {author}
  {\bibfnamefont {D.~S.}\ \bibnamefont {Jin}}, \ and\ \bibinfo {author}
  {\bibfnamefont {E.~A.}\ \bibnamefont {Cornell}},\ }\href@noop {} {\bibfield
  {journal} {\bibinfo  {journal} {Phys. Rev. Lett.}\ }\textbf {\bibinfo
  {volume} {119}},\ \bibinfo {pages} {143401} (\bibinfo {year}
  {2017})}\BibitemShut {NoStop}%
\bibitem [{\citenamefont {Jochim}\ \emph {et~al.}(2003)\citenamefont {Jochim},
  \citenamefont {Bartenstein}, \citenamefont {Altmeyer}, \citenamefont {Hendl},
  \citenamefont {Riedl}, \citenamefont {Chin}, \citenamefont {Denschlag},\ and\
  \citenamefont {Grimm}}]{jochim2003bose}%
  \BibitemOpen
  \bibfield  {author} {\bibinfo {author} {\bibfnamefont {S.}~\bibnamefont
  {Jochim}}, \bibinfo {author} {\bibfnamefont {M.}~\bibnamefont {Bartenstein}},
  \bibinfo {author} {\bibfnamefont {A.}~\bibnamefont {Altmeyer}}, \bibinfo
  {author} {\bibfnamefont {G.}~\bibnamefont {Hendl}}, \bibinfo {author}
  {\bibfnamefont {S.}~\bibnamefont {Riedl}}, \bibinfo {author} {\bibfnamefont
  {C.}~\bibnamefont {Chin}}, \bibinfo {author} {\bibfnamefont {J.~H.}\
  \bibnamefont {Denschlag}}, \ and\ \bibinfo {author} {\bibfnamefont
  {R.}~\bibnamefont {Grimm}},\ }\href@noop {} {\bibfield  {journal} {\bibinfo
  {journal} {Science}\ }\textbf {\bibinfo {volume} {302}},\ \bibinfo {pages}
  {2101} (\bibinfo {year} {2003})}\BibitemShut {NoStop}%
\bibitem [{\citenamefont {Greiner}\ \emph {et~al.}(2003)\citenamefont
  {Greiner}, \citenamefont {Regal},\ and\ \citenamefont
  {Jin}}]{greiner2003emergence}%
  \BibitemOpen
  \bibfield  {author} {\bibinfo {author} {\bibfnamefont {M.}~\bibnamefont
  {Greiner}}, \bibinfo {author} {\bibfnamefont {C.~A.}\ \bibnamefont {Regal}},
  \ and\ \bibinfo {author} {\bibfnamefont {D.~S.}\ \bibnamefont {Jin}},\
  }\href@noop {} {\bibfield  {journal} {\bibinfo  {journal} {Nature}\ }\textbf
  {\bibinfo {volume} {426}},\ \bibinfo {pages} {537} (\bibinfo {year}
  {2003})}\BibitemShut {NoStop}%
\bibitem [{\citenamefont {Herbig}\ \emph {et~al.}(2003)\citenamefont {Herbig},
  \citenamefont {Kraemer}, \citenamefont {Mark}, \citenamefont {Weber},
  \citenamefont {Chin}, \citenamefont {N{\"a}gerl},\ and\ \citenamefont
  {Grimm}}]{herbig2003preparation}%
  \BibitemOpen
  \bibfield  {author} {\bibinfo {author} {\bibfnamefont {J.}~\bibnamefont
  {Herbig}}, \bibinfo {author} {\bibfnamefont {T.}~\bibnamefont {Kraemer}},
  \bibinfo {author} {\bibfnamefont {M.}~\bibnamefont {Mark}}, \bibinfo {author}
  {\bibfnamefont {T.}~\bibnamefont {Weber}}, \bibinfo {author} {\bibfnamefont
  {C.}~\bibnamefont {Chin}}, \bibinfo {author} {\bibfnamefont {H.-C.}\
  \bibnamefont {N{\"a}gerl}}, \ and\ \bibinfo {author} {\bibfnamefont
  {R.}~\bibnamefont {Grimm}},\ }\href@noop {} {\bibfield  {journal} {\bibinfo
  {journal} {Science}\ }\textbf {\bibinfo {volume} {301}},\ \bibinfo {pages}
  {1510} (\bibinfo {year} {2003})}\BibitemShut {NoStop}%
\bibitem [{\citenamefont {Spiegelhalder}\ \emph {et~al.}(2010)\citenamefont
  {Spiegelhalder}, \citenamefont {Trenkwalder}, \citenamefont {Naik},
  \citenamefont {Kerner}, \citenamefont {Wille}, \citenamefont {Hendl},
  \citenamefont {Schreck},\ and\ \citenamefont {Grimm}}]{spiegelhalder2010all}%
  \BibitemOpen
  \bibfield  {author} {\bibinfo {author} {\bibfnamefont {F.}~\bibnamefont
  {Spiegelhalder}}, \bibinfo {author} {\bibfnamefont {A.}~\bibnamefont
  {Trenkwalder}}, \bibinfo {author} {\bibfnamefont {D.}~\bibnamefont {Naik}},
  \bibinfo {author} {\bibfnamefont {G.}~\bibnamefont {Kerner}}, \bibinfo
  {author} {\bibfnamefont {E.}~\bibnamefont {Wille}}, \bibinfo {author}
  {\bibfnamefont {G.}~\bibnamefont {Hendl}}, \bibinfo {author} {\bibfnamefont
  {F.}~\bibnamefont {Schreck}}, \ and\ \bibinfo {author} {\bibfnamefont
  {R.}~\bibnamefont {Grimm}},\ }\href@noop {} {\bibfield  {journal} {\bibinfo
  {journal} {Phys. Rev. A}\ }\textbf {\bibinfo {volume} {81}},\ \bibinfo
  {pages} {043637} (\bibinfo {year} {2010})}\BibitemShut {NoStop}%
\bibitem [{\citenamefont {Ospelkaus}\ \emph {et~al.}(2006)\citenamefont
  {Ospelkaus}, \citenamefont {Ospelkaus}, \citenamefont {Humbert},
  \citenamefont {Ernst}, \citenamefont {Sengstock},\ and\ \citenamefont
  {Bongs}}]{ospelkaus2006ultracold}%
  \BibitemOpen
  \bibfield  {author} {\bibinfo {author} {\bibfnamefont {C.}~\bibnamefont
  {Ospelkaus}}, \bibinfo {author} {\bibfnamefont {S.}~\bibnamefont
  {Ospelkaus}}, \bibinfo {author} {\bibfnamefont {L.}~\bibnamefont {Humbert}},
  \bibinfo {author} {\bibfnamefont {P.}~\bibnamefont {Ernst}}, \bibinfo
  {author} {\bibfnamefont {K.}~\bibnamefont {Sengstock}}, \ and\ \bibinfo
  {author} {\bibfnamefont {K.}~\bibnamefont {Bongs}},\ }\href@noop {}
  {\bibfield  {journal} {\bibinfo  {journal} {Phys. Rev. Lett.}\ }\textbf
  {\bibinfo {volume} {97}},\ \bibinfo {pages} {120402} (\bibinfo {year}
  {2006})}\BibitemShut {NoStop}%
\bibitem [{\citenamefont {Zirbel}\ \emph {et~al.}(2008)\citenamefont {Zirbel},
  \citenamefont {Ni}, \citenamefont {Ospelkaus}, \citenamefont {Nicholson},
  \citenamefont {Olsen}, \citenamefont {Julienne}, \citenamefont {Wieman},
  \citenamefont {Ye},\ and\ \citenamefont {Jin}}]{zirbel2008heteronuclear}%
  \BibitemOpen
  \bibfield  {author} {\bibinfo {author} {\bibfnamefont {J.}~\bibnamefont
  {Zirbel}}, \bibinfo {author} {\bibfnamefont {K.-K.}\ \bibnamefont {Ni}},
  \bibinfo {author} {\bibfnamefont {S.}~\bibnamefont {Ospelkaus}}, \bibinfo
  {author} {\bibfnamefont {T.}~\bibnamefont {Nicholson}}, \bibinfo {author}
  {\bibfnamefont {M.}~\bibnamefont {Olsen}}, \bibinfo {author} {\bibfnamefont
  {P.}~\bibnamefont {Julienne}}, \bibinfo {author} {\bibfnamefont
  {C.}~\bibnamefont {Wieman}}, \bibinfo {author} {\bibfnamefont
  {J.}~\bibnamefont {Ye}}, \ and\ \bibinfo {author} {\bibfnamefont
  {D.}~\bibnamefont {Jin}},\ }\href@noop {} {\bibfield  {journal} {\bibinfo
  {journal} {Phys. Rev. A}\ }\textbf {\bibinfo {volume} {78}},\ \bibinfo
  {pages} {013416} (\bibinfo {year} {2008})}\BibitemShut {NoStop}%
\bibitem [{\citenamefont {D{\"u}rr}\ \emph {et~al.}(2004)\citenamefont
  {D{\"u}rr}, \citenamefont {Volz},\ and\ \citenamefont
  {Rempe}}]{durr2004dissociation}%
  \BibitemOpen
  \bibfield  {author} {\bibinfo {author} {\bibfnamefont {S.}~\bibnamefont
  {D{\"u}rr}}, \bibinfo {author} {\bibfnamefont {T.}~\bibnamefont {Volz}}, \
  and\ \bibinfo {author} {\bibfnamefont {G.}~\bibnamefont {Rempe}},\
  }\href@noop {} {\bibfield  {journal} {\bibinfo  {journal} {Phys. Rev. A}\
  }\textbf {\bibinfo {volume} {70}},\ \bibinfo {pages} {031601} (\bibinfo
  {year} {2004})}\BibitemShut {NoStop}%
\bibitem [{\citenamefont {Wang}\ \emph {et~al.}(2019)\citenamefont {Wang},
  \citenamefont {Ye}, \citenamefont {Guo}, \citenamefont {Blume},\ and\
  \citenamefont {Wang}}]{wang2019observation}%
  \BibitemOpen
  \bibfield  {author} {\bibinfo {author} {\bibfnamefont {F.}~\bibnamefont
  {Wang}}, \bibinfo {author} {\bibfnamefont {X.}~\bibnamefont {Ye}}, \bibinfo
  {author} {\bibfnamefont {M.}~\bibnamefont {Guo}}, \bibinfo {author}
  {\bibfnamefont {D.}~\bibnamefont {Blume}}, \ and\ \bibinfo {author}
  {\bibfnamefont {D.}~\bibnamefont {Wang}},\ }\href@noop {} {\bibfield
  {journal} {\bibinfo  {journal} {Phys. Rev. A}\ }\textbf {\bibinfo {volume}
  {100}},\ \bibinfo {pages} {042706} (\bibinfo {year} {2019})}\BibitemShut
  {NoStop}%
\bibitem [{\citenamefont {Wang}\ \emph {et~al.}(2015)\citenamefont {Wang},
  \citenamefont {He}, \citenamefont {Li}, \citenamefont {Zhu}, \citenamefont
  {Chen},\ and\ \citenamefont {Wang}}]{wang2015formation}%
  \BibitemOpen
  \bibfield  {author} {\bibinfo {author} {\bibfnamefont {F.}~\bibnamefont
  {Wang}}, \bibinfo {author} {\bibfnamefont {X.}~\bibnamefont {He}}, \bibinfo
  {author} {\bibfnamefont {X.}~\bibnamefont {Li}}, \bibinfo {author}
  {\bibfnamefont {B.}~\bibnamefont {Zhu}}, \bibinfo {author} {\bibfnamefont
  {J.}~\bibnamefont {Chen}}, \ and\ \bibinfo {author} {\bibfnamefont
  {D.}~\bibnamefont {Wang}},\ }\href@noop {} {\bibfield  {journal} {\bibinfo
  {journal} {New J. Phys.}\ }\textbf {\bibinfo {volume} {17}},\ \bibinfo
  {pages} {035003} (\bibinfo {year} {2015})}\BibitemShut {NoStop}%
\bibitem [{\citenamefont {Safronova}\ \emph {et~al.}(2006)\citenamefont
  {Safronova}, \citenamefont {Arora},\ and\ \citenamefont
  {Clark}}]{safronova2006frequency}%
  \BibitemOpen
  \bibfield  {author} {\bibinfo {author} {\bibfnamefont {M.}~\bibnamefont
  {Safronova}}, \bibinfo {author} {\bibfnamefont {B.}~\bibnamefont {Arora}}, \
  and\ \bibinfo {author} {\bibfnamefont {C.~W.}\ \bibnamefont {Clark}},\
  }\href@noop {} {\bibfield  {journal} {\bibinfo  {journal} {Phys. Rev. A}\
  }\textbf {\bibinfo {volume} {73}},\ \bibinfo {pages} {022505} (\bibinfo
  {year} {2006})}\BibitemShut {NoStop}%
\bibitem [{\citenamefont {Schünemann}\ \emph {et~al.}(1999)\citenamefont
  {Schünemann}, \citenamefont {Engler}, \citenamefont {Grimm}, \citenamefont
  {Weidemüller},\ and\ \citenamefont {Zielonkowski}}]{Schunemann1999offset}%
  \BibitemOpen
  \bibfield  {author} {\bibinfo {author} {\bibfnamefont {U.}~\bibnamefont
  {Schünemann}}, \bibinfo {author} {\bibfnamefont {H.}~\bibnamefont {Engler}},
  \bibinfo {author} {\bibfnamefont {R.}~\bibnamefont {Grimm}}, \bibinfo
  {author} {\bibfnamefont {M.}~\bibnamefont {Weidemüller}}, \ and\ \bibinfo
  {author} {\bibfnamefont {M.}~\bibnamefont {Zielonkowski}},\ }\href {\doibase
  10.1063/1.1149573} {\bibfield  {journal} {\bibinfo  {journal} {Rev. Sci.
  Instrum.}\ }\textbf {\bibinfo {volume} {70}},\ \bibinfo {pages} {242}
  (\bibinfo {year} {1999})}\BibitemShut {NoStop}%
\bibitem [{\citenamefont {McGuyer}\ \emph {et~al.}(2013)\citenamefont
  {McGuyer}, \citenamefont {Osborn}, \citenamefont {McDonald}, \citenamefont
  {Reinaudi}, \citenamefont {Skomorowski}, \citenamefont {Moszynski},\ and\
  \citenamefont {Zelevinsky}}]{mcguyer2013nonadiabatic}%
  \BibitemOpen
  \bibfield  {author} {\bibinfo {author} {\bibfnamefont {B.}~\bibnamefont
  {McGuyer}}, \bibinfo {author} {\bibfnamefont {C.}~\bibnamefont {Osborn}},
  \bibinfo {author} {\bibfnamefont {M.}~\bibnamefont {McDonald}}, \bibinfo
  {author} {\bibfnamefont {G.}~\bibnamefont {Reinaudi}}, \bibinfo {author}
  {\bibfnamefont {W.}~\bibnamefont {Skomorowski}}, \bibinfo {author}
  {\bibfnamefont {R.}~\bibnamefont {Moszynski}}, \ and\ \bibinfo {author}
  {\bibfnamefont {T.}~\bibnamefont {Zelevinsky}},\ }\href@noop {} {\bibfield
  {journal} {\bibinfo  {journal} {Phys. Rev. Lett.}\ }\textbf {\bibinfo
  {volume} {111}},\ \bibinfo {pages} {243003} (\bibinfo {year}
  {2013})}\BibitemShut {NoStop}%
\bibitem [{\citenamefont {Ye}\ \emph {et~al.}(2018)\citenamefont {Ye},
  \citenamefont {Guo}, \citenamefont {Gonz{\'a}lez-Mart{\'\i}nez},
  \citenamefont {Qu{\'e}m{\'e}ner},\ and\ \citenamefont
  {Wang}}]{ye2018collisions}%
  \BibitemOpen
  \bibfield  {author} {\bibinfo {author} {\bibfnamefont {X.}~\bibnamefont
  {Ye}}, \bibinfo {author} {\bibfnamefont {M.}~\bibnamefont {Guo}}, \bibinfo
  {author} {\bibfnamefont {M.~L.}\ \bibnamefont {Gonz{\'a}lez-Mart{\'\i}nez}},
  \bibinfo {author} {\bibfnamefont {G.}~\bibnamefont {Qu{\'e}m{\'e}ner}}, \
  and\ \bibinfo {author} {\bibfnamefont {D.}~\bibnamefont {Wang}},\ }\href@noop
  {} {\bibfield  {journal} {\bibinfo  {journal} {Sci. Adv.}\ }\textbf {\bibinfo
  {volume} {4}},\ \bibinfo {pages} {eaaq0083} (\bibinfo {year}
  {2018})}\BibitemShut {NoStop}%
\bibitem [{\citenamefont {Guo}\ \emph {et~al.}(2018)\citenamefont {Guo},
  \citenamefont {Ye}, \citenamefont {He}, \citenamefont
  {Gonz\'alez-Mart\'{\i}nez}, \citenamefont {Vexiau}, \citenamefont
  {Qu\'em\'ener},\ and\ \citenamefont {Wang}}]{Guo2018}%
  \BibitemOpen
  \bibfield  {author} {\bibinfo {author} {\bibfnamefont {M.}~\bibnamefont
  {Guo}}, \bibinfo {author} {\bibfnamefont {X.}~\bibnamefont {Ye}}, \bibinfo
  {author} {\bibfnamefont {J.}~\bibnamefont {He}}, \bibinfo {author}
  {\bibfnamefont {M.~L.}\ \bibnamefont {Gonz\'alez-Mart\'{\i}nez}}, \bibinfo
  {author} {\bibfnamefont {R.}~\bibnamefont {Vexiau}}, \bibinfo {author}
  {\bibfnamefont {G.}~\bibnamefont {Qu\'em\'ener}}, \ and\ \bibinfo {author}
  {\bibfnamefont {D.}~\bibnamefont {Wang}},\ }\href {\doibase
  10.1103/PhysRevX.8.041044} {\bibfield  {journal} {\bibinfo  {journal} {Phys.
  Rev. X}\ }\textbf {\bibinfo {volume} {8}},\ \bibinfo {pages} {041044}
  (\bibinfo {year} {2018})}\BibitemShut {NoStop}%
\end{thebibliography}

\providecommand{\noopsort}[1]{}\providecommand{\singleletter}[1]{#1}%

\end{document}